# Distributions of $H_2O$ and $CO_2$ ices on Ariel, Umbriel, Titania, and Oberon from IRTF/SpeX observations

W.M. Grundy[1,2], L.A. Young[1,3], J.R. Spencer[3], R.E. Johnson[4], E.F. Young[1,3], and M.W. Buie[2]

1. Visiting/remote observer at the Infrared Telescope Facility, which is operated by the University of Hawaii under contract from NASA.
2. Lowell Observatory, 1400 W. Mars Hill Rd., Flagstaff AZ 86001.
3. Southwest Research Institute, 1050 Walnut St., Boulder CO 80302.
4. Univ. of Virginia Dept. Engineering Physics, 116 Engineer's Way, Charlottesville VA 22904.

## Abstract

We present 0.8 to 2.4 µm spectral observations of uranian satellites, obtained at IRTF/SpeX on 17 nights during 2001-2005. The spectra reveal for the first time the presence of $CO_2$ ice on the surfaces of Umbriel and Titania, by means of 3 narrow absorption bands near 2 µm. Several additional, weaker $CO_2$ ice absorptions have also been detected. No $CO_2$ absorption is seen in Oberon spectra, and the strengths of the $CO_2$ ice bands decline with planetocentric distance from Ariel through Titania. We use the $CO_2$ absorptions to map the longitudinal distribution of $CO_2$ ice on Ariel, Umbriel, and Titania, showing that it is most abundant on their trailing hemispheres. We also examine $H_2O$ ice absorptions in the spectra, finding deeper $H_2O$ bands on the leading hemispheres of Ariel, Umbriel, and Titania, but the opposite pattern on Oberon. Potential mechanisms to produce the observed longitudinal and planetocentric distributions of the two ices are considered.

# 1. Introduction

The discovery of $CO_2$ ice on the surface of the uranian satellite Ariel (Grundy et al. 2003) raised a number of questions regarding its origin and ultimate fate. Is it a relic of Ariel's primordial volatile inventory? Was it delivered by more recent impactors? Or is it being produced *in situ* in chemical reactions enabled by energetic radiation? How long does exposed $CO_2$ ice survive on the satellite's surface, and what processes act on it? How stable is it against sublimation and sputtering? Does it participate in radiolytic chemistry, such as envisioned by Delitsky and Lane (1997, 1998), Hudson and Moore (2001), and Johnson (2001)?

Questions such as these could potentially be addressed by means of a survey of $CO_2$ ice as a function of longitude and planetocentric distance among the major uranian satellites, the basic physical properties of which are listed in Table 1. The results of such a survey of the four largest satellites are presented in this paper.

**Table 1.** Major satellites of Uranus.

| Satellite | Distance from Uranus ($10^5$ km) | Orbital Period (days) | Radius[a] (km) | Mass[b] ($10^{23}$ g) | Surface gravity (cm s$^{-2}$) | Visual full disk albedo[c] |
|---|---|---|---|---|---|---|
| Miranda | 1.30 | 1.41 | 236 | 0.66 | 7.9 | 0.307 |
| Ariel | 1.91 | 2.52 | 579 | 13.5 | 26.9 | 0.350 |
| Umbriel | 2.66 | 4.14 | 585 | 11.7 | 22.9 | 0.189 |
| Titania | 4.36 | 8.71 | 789 | 35.3 | 37.8 | 0.232 |
| Oberon | 5.81 | 13.46 | 761 | 30.1 | 34.7 | 0.205 |

[a]Radii are from Thomas (1988).

[b]Masses are from Jacobson et al. (1992).

[c]Full disk albedos at 0.55 µm and 1° phase are from Karkoschka (1997) Table II.

# 2. Observations and Reduction

We obtained infrared spectroscopy of uranian satellites on seventeen nights during 2001-2005 at NASA's Infrared Telescope Facility (IRTF) on Mauna Kea, as tabulated in Table 2. The data were collected using the short cross-dispersed mode of the SpeX spectrograph (Rayner et al. 1998, 2003), which covers the 0.8 to 2.4 µm wavelength range in five spectral orders, recorded simultaneously on a 1024 × 1024 InSb array. Uranian satellite observations were done at an average airmass of 1.26. Maximum airmasses were 1.61, 1.77, 1.57, and 1.46 for Ariel, Umbriel, Titania, and Oberon, respectively. When observing at higher airmasses, we used an image rotator to keep SpeX's slit oriented near the parallactic angle on the sky-plane, to minimize wavelength-dependent slit losses from differential atmospheric refraction. During 2001-2002 we used the 0.5 arcsec slit, and during 2003-2005 we used the 0.3 arcsec slit, resulting in measured spectral resolutions ($\lambda/\Delta\lambda$ for unresolved OH airglow emission lines) between 1300 and 1400 and between 1600 and 1700, respectively.

All spectra were collected as pairs, with the telescope being offset along the slit between two locations, referred to as A and B beams. Pairs of A and B spectral frames were subtracted prior to extraction using the Horne (1986) optimal extraction algorithm as implemented by M.W. Buie et al. at Lowell Observatory (e.g., Buie and Grundy 2000).

**Table 2.** Circumstances of observations

| UT date of observation mid-time | Sky conditions and H band image size | Sub-observer longitude (°) | Sub-observer latitude (°) | Phase angle (°) | Total integration (min) |
|---|---|---|---|---|---|
| **Ariel** | | | | | |
| 2001/07/05.59 | Cirrus, 1.1" | 233.8 | -23.1 | 1.91 | 50 |
| 2001/07/08.61 | Clear, 1.1" | 304.8 | -23.2 | 1.79 | 48 |
| 2002/07/16.55 | Partly cloudy, 0.7" | 294.8 | -19.3 | 1.65 | 140 |
| 2002/07/17.56 | Partly cloudy, 0.8" | 79.8 | -19.4 | 1.60 | 108 |
| 2003/08/05.50 | Clear, 0.6" | 200.0 | -15.9 | 0.95 | 84 |
| 2003/08/09.51 | Clear, 0.6" | 53.6 | -16.0 | 0.75 | 156 |
| 2003/09/07.40 | Clear, 0.6" | 219.8 | -17.2 | 0.70 | 90 |
| 2003/10/04.24 | Clear, 0.6" | 93.5 | -18.1 | 1.88 | 108 |
| 2003/10/08.33 | Clear, 0.9" | 316.6 | -18.2 | 2.03 | 132 |
| 2004/07/15.50 | Clear, 0.6" | 159.9 | -11.1 | 1.99 | 112 |
| **Umbriel** | | | | | |
| 2001/07/07.59 | Cirrus, 1.0" | 219.8 | -23.0 | 1.83 | 52 |
| 2004/07/05.59 | Thin cirrus, 0.8" | 216.2 | -10.9 | 2.32 | 80 |
| 2004/07/16.60 | Thin cirrus, 0.8" | 92.1 | -11.1 | 1.95 | 74 |
| 2004/07/27.48 | Clear, 0.6" | 317.6 | -11.4 | 1.51 | 184 |
| 2005/09/18.39 | Clear, 0.6" | 159.9 | - 9.4 | 0.85 | 196 |
| **Titania** | | | | | |
| 2001/07/06.56 | Cirrus, 0.9" | 237.0 | -23.0 | 1.87 | 56 |
| 2001/07/07.55 | Cirrus, 1.0" | 277.8 | -23.0 | 1.83 | 36 |
| 2003/10/08.24 | Clear, 0.9" | 98.0 | -18.1 | 2.03 | 64 |
| 2004/07/15.60 | Clear, 0.6" | 213.9 | -11.1 | 1.98 | 64 |
| 2005/10/13.38 | Clear, 0.6" | 299.6 | -10.2 | 1.93 | 106 |
| **Oberon** | | | | | |
| 2001/07/06.60 | Cirrus, 0.9" | 164.0 | -23.0 | 1.87 | 32 |
| 2001/07/08.55 | Clear, 1.1" | 216.2 | -23.1 | 1.79 | 48 |
| 2005/10/13.27 | Clear, 0.6" | 110.6 | -10.2 | 1.93 | 124 |

Uranian satellite observations were interspersed between observations of nearby solar analog reference stars HD 210377 and HD 214572. We also observed more distant, but better-known solar analogs 16 Cyg B, BS 5968, BS 6060, HD 219018, and SA 112-1333. Those observations showed that HD 214572 is a reasonable solar analog in this spectral range, and that HD 210377 was also usable as a solar analog after correcting for its effective temperature, which is a few hundred K cooler than the sun. From the solar analog observations obtained each night we computed telluric extinction and corrected the star and satellite observations to a common airmass. We then divided the airmass-corrected satellite spectra by the airmass-corrected solar analog spectra. This operation eliminated most instrumental, stellar, and telluric spectral features, and produced spectra proportional to the satellite disk-integrated albedos. Cancellation of telluric features was sometimes imperfect near 1.4 and 1.9 µm, where strong and narrow telluric $H_2O$ vapor absorptions make sky transparency especially variable in time. Additional spurious features originate in scattered light from Uranus, such as at 0.94 µm. These could be either positive or negative, depending whether Uranus was closer to the object or sky beam position, and only affected Umbriel and Ariel spectra for longitudes near 0 or 180°. Examples of final, normalized albedo spectra are shown in Fig. 1.

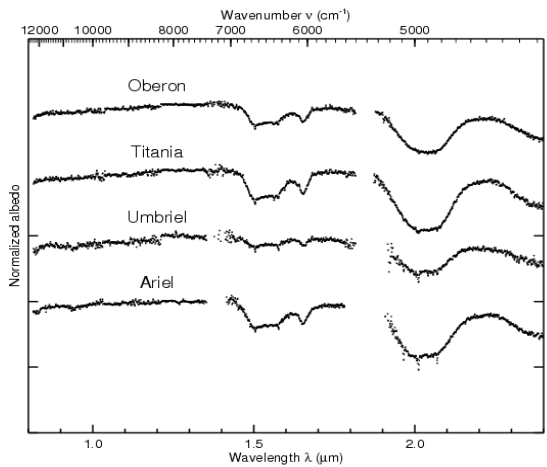

**Figure 1:** Example spectra of uranian satellites, normalized at *J* band and offset upward with zero levels indicated by tick marks on the ordinate axis. Gaps around 1.4 and 1.85 µm coincide with high atmospheric opacity. The Ariel spectrum is an average of all spectra with subsolar longitudes between 210° and 330°. The Umbriel spectrum is from 2005/09/18 UT and the Titania and Oberon spectra are both from 2005/10/13 UT.

Wavelength calibration was derived from telluric airglow emission lines, primarily those of OH, extracted from the satellite frames. The dispersion ranges from 0.00023 to 0.00054 µm per pixel from the shortest to longest wavelengths. The wavelength uncertainty is no more than a pixel in regions having abundant sky emission lines, which include the wavelengths where the satellites exhibit absorption features analyzed in this paper.

Our spectra could not be photometrically calibrated, since variable slit losses from different tracking time scales combine with variable focus and seeing to undermine photometric fidelity of comparisons between the faint satellites and the much brighter solar analog stars. However, other sources of photometry exist for the uranian satellites in our wavelength range (e.g., Karkoschka 1997) and these could be used to scale the spectra to an albedo scale, enabling the use of quantitative radiative transfer models.

## 3. Analysis

The spectra of all four satellites in Fig. 1 look broadly similar to one another. A gentle reddish spectral slope extends from 0.8 to about 1.4 µm, followed by a water ice absorption complex between 1.45 and 1.7 µm, and two more $H_2O$ ice absorptions between 1.9 and 2.2 µm and from 2.25 µm to the limit of our spectral coverage at 2.4 µm. The $H_2O$ ice bands are weakest in the spectrum of Umbriel, the satellite with the lowest albedo of the four (Karkoschka 1997). Lower albedos and weaker $H_2O$ bands typically go together, since the same dark materials (probably carbonaceous, in the case of uranian satellites) which suppress albedos are also effective at masking $H_2O$ ice absorptions. Although this dark material has a profound effect on the satellites' albedos, $H_2O$ ice could still be the most abundant surface material on all four satellites, considering that thermal segregation of $H_2O$ ice, as seen on the jovian satellites, is improbable at the low surface temperatures of the uranian satellites (Spencer 1987). Impact gardening probably leads to intimate mixing between dark and icy components, enabling even small quantities of dark material to significantly suppress surface albedos (Clark 1981).

## 3.1. Water ice

The shapes of the H$_2$O ice absorption bands and their depths relative to one another, especially between 1.50 and 1.57 µm and around 1.65 µm, confirm that most of the H$_2$O ice on all four satellites is crystalline at the ~mm depths sampled by these wavelengths (e.g., Grundy et al. 1999; Hansen and McCord 2004). Amorphous ice lacks 1.57 and 1.65 µm absorption bands and exhibits somewhat different shapes for the 1.5 and 2 µm bands (Grundy and Schmitt 1998; Schmitt et al. 1998). The H$_2$O features in the satellite spectra are consistent with simple Hapke models (e.g., Hapke 1993) having no amorphous ice at all, although we can include up to 10 to 20% amorphous ice before the discrepancies become conspicuous.

This predominance of crystalline ice is not surprising, since it is the more thermodynamically stable, lower energy phase. Crystalline H$_2$O has also been reported on other outer solar system surfaces at comparable and greater heliocentric distances, both icy satellites and trans-neptunian objects. In fact, everywhere that the phase of H$_2$O ice on these objects' surfaces has been identified, it is crystalline (e.g., Buie and Grundy 2000; Bauer et al. 2002; Grundy and Young 2004; Jewitt and Luu 2004).

Under certain circumstances, energetic charged particles or UV photons can disrupt the crystal structure of ice (e.g., Kouchi and Kuroda 1990; Johnson 2000; Moore and Hudson 1992; Johnson and Quickenden 1997), but evidently this effect is so inefficient that ice on the uranian satellites is not amorphous at mm depths. Nature does tend to find its way to thermodynamically favored states such as crystallinity, in the case of H$_2$O ice. Thermal recrystallization goes as $e^{-E_A/kT}$, where $E_A$ is an activation energy, $k$ is the Boltzmann constant, and $T$ is the temperature. Laboratory studies of crystallization rates of pure H$_2$O ice at much warmer temperatures (Delitsky and Lane 1998; Jenniskens et al. 1998), if extrapolated over many orders of magnitude, suggest that thermally activated crystallization could be extremely slow at uranian satellite surface temperatures. However, even if the low thermal crystallization rates implied by that extrapolation are correct, the presence of impurities in the ice, or of non-thermal energy inputs from less energetic charged particles, photons, or even micrometeorites may help overcome thermal limits (e.g., Brown et al. 1978). Sputtering might also remove H$_2$O more efficiently than it is amorphized. Consequently, we do not subscribe to the idea proposed by Bauer et al. 2002 and by Jewitt and Luu (2004) that the existence of crystalline ice on the surfaces of these bodies implies recent heating episodes.

The H$_2$O ice absorption bands of jovian and saturnian satellites tend to be deeper on their leading hemispheres (e.g., Clark et al. 1984; Grundy et al. 1999; Cruikshank et al. 2005). Similar behavior is exhibited by the uranian satellites, except for Oberon, as shown in Fig. 2. Integrated areas of the 1.5 µm water ice absorption complex were computed by normalizing each spectrum to a line fitted to continuum wavelengths on either side of the absorption band (1.29 to 1.32 and 1.72 to 1.75 µm, in this example), then integrating one minus the normalized spectrum over the band interval (1.42 to 1.72 µm). Ariel, with the most complete longitudinal sampling, shows a sinusoidal pattern of variation of its integrated H$_2$O ice band, with maximum H$_2$O absorption coinciding with the leading hemisphere (90° longitude in the left-handed IAU system). Fewer longitudes were observed for the other satellites, but Umbriel and Titania both show more H$_2$O absorption on their leading hemispheres. The trend for Oberon appears to be reversed, but confirming observations are needed.

Magnetospheric charged particle bombardment can drive sputtering as well as radiolytic destruction of H$_2$O ice, and could be responsible for the observed leading-trailing asymmetries. The magnetic field of Uranus ro-tates rigidly, at the same frequency as Uranus spins on its axis, with a period of 17.9 hours. The satellites orbit more slowly (see Table 1), so the magnetosphere overtakes them from behind and its charged particles preferentially strike their trailing hemispheres (Cheng et al. 1986; Lanzerotti et al. 1987). Magnetospheric charged particles spiral along the magnetic field lines at gyroradii inversely proportional to the strength of the magnetic field. The dipole component of the field diminishes as the inverse cube of the distance, so gyroradii grow as the cube of the distance from Uranus. For protons, the dominant ions in the uranian magnetosphere (Bridge et al. 1986), the energies at which gyroradii match the radii of the satellites range from ~100 keV for Miranda down to <1 keV for Titania, spanning the 'dominant' ~10 keV energy range for magnetospheric protons (Johnson 2000). That these protons have gyroradii larger than the radii of the satellites in the outer parts of the system suggests that outer satellites should receive more isotropicially distributed proton bombardment. This tendency is moderated by the fact that the net plasma flow past the satellites, arising from the greater speed of the uranian magnetic field relative to the Keplerian motion of the satellites, is faster further from Uranus. This net plasma flow increases from 6 up to 39 km $sec^{-1}$ from Miranda to Titania. However, these speeds correspond to proton energies only 0.2 to 8 eV, far below 10 keV energies. Although 10 keV protons circle field lines in orbits larger than the radii of satellites from Umbriel on out, the orbits are executed relatively quickly compared with the rate at which the magnetic field sweeps past the satellites. Orbital periods for protons are inversely proportional to magnetic field strength, being 0.3, 1, 3, and 13 seconds, from Miranda through Titania. Combining these gyroperiods with net plasma flow rates, each successive proton orbit advances by 2, 14, 60, and 500 km, relative to satellites Miranda through Titania. Only at Titania and beyond does a 10 keV proton advance far enough between successive loops to begin to have much chance of looping around and hitting the satellite's leading hemisphere. Computing the actual three dimensional spatial distribution of impacts on a satellite's surface as a function of proton energy is beyond the scope of this paper. It depends on the substantial tilt and offset of the dipole field of Uranus as well as the distribution of particle pitch angles, which could vary seasonally. However, the simple arguments presented here suggest that particles in the 10 keV range will mostly hit trailing hemispheres of satellites, and can only influence the leading hemispheres of the more distant ones. Additionally, magnetospheric plasma densities decline rapidly with planetocentric distance, more rapidly than is compensated by the increase in net plasma flow relative to satellite orbital motion (or at least they did during the Voyager 2 encounter in 1986, near the time of southern summer solstice, e.g., Bridge et al. 1986; Cheng et al. 1991). Overall magnetospheric plasma sputtering rates should thus be lower on the more distant satellites. For both of these reasons, we would expect magnetospheric sputtering to produce diminishing leading-trailing contrasts with planetocentric distance. This expectation is consistent with results from sinusoidal fits to the observed longitudinal variation in $H_2O$ bands, which give fractional variations $93 \pm 4\%$, $69 \pm 47\%$, and $34 \pm 13\%$ for Ariel, Umbriel, and Titania, respectively.

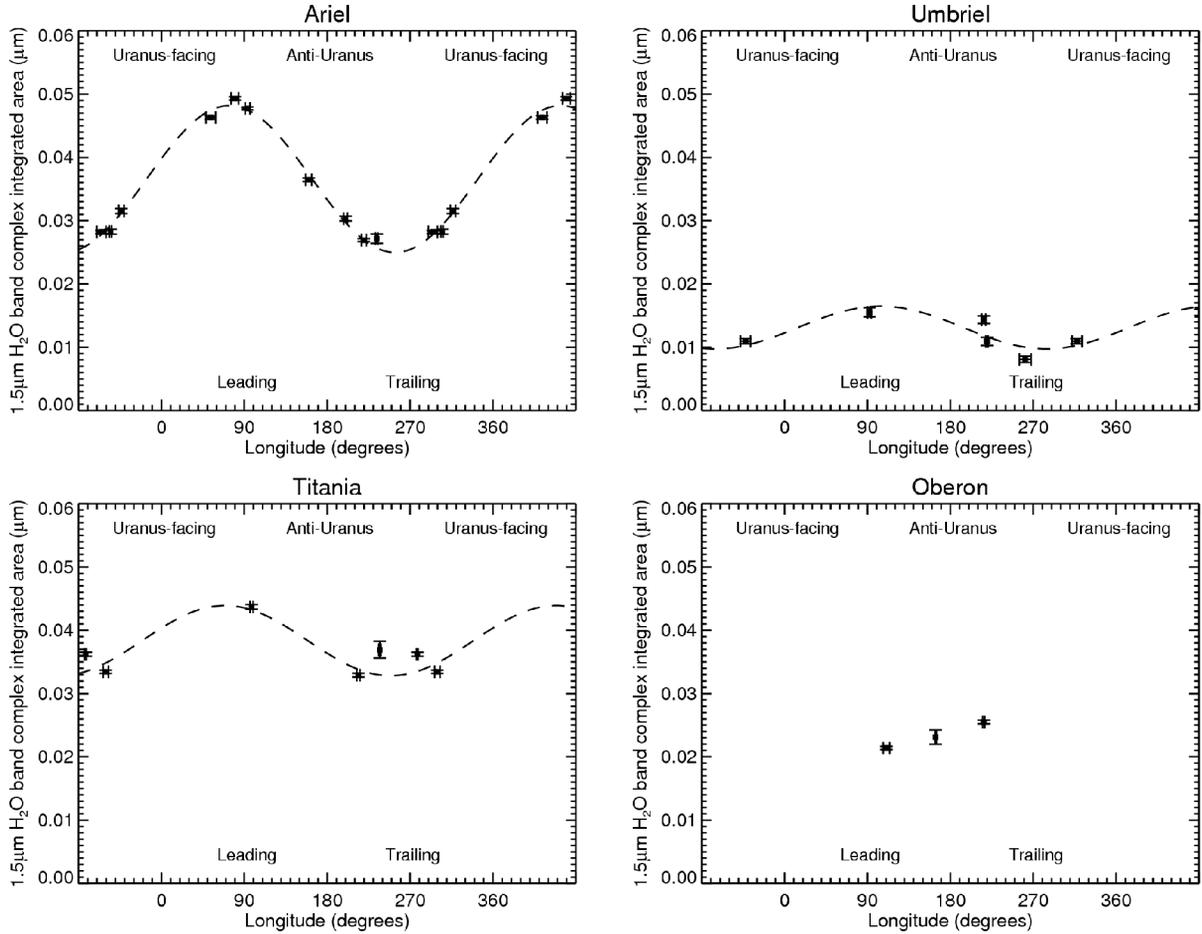

**Figure 2:** Variation of integrated area of the 1.5 μm $H_2O$ ice band complex as a function of sub-solar/sub-viewer longitude, showing large cyclical variations with more $H_2O$ ice absorption on the leading hemispheres, at least for satellites with better longitude coverage, where sine fits to the data are plotted as dashed curves. Data are duplicated outside the 0° to 360° interval to better show the periodic trends.

Another mechanism capable of producing leading-trailing asymmetries in the satellites' $H_2O$ ice bands is magnetospheric delivery of carbonaceous dust particles. Dark particles from the uranian rings which become electrically charged will experience a Lorentz force from the magnetic field lines sweeping past them. This force will tend to accelerate them such that they spiral slowly outward, eventually culminating in low speed collisions onto the trailing hemispheres of the satellites, primarily the inner ones, so this mechanism could also contribute to the observed decline of leading-trailing spectral contrasts with planetocentric distance. The addition of these dark particles to the satellites' trailing hemispheres would suppress their albedos and their ice absorption bands, by reducing the fraction of photons able to escape from the surface at all wavelengths, as opposed to just the wavelengths at which ice absorbs. A possible way to establish limits on the supply of carbonaceous grains via this mechanism could involve comparison of visible wavelength albedos between leading and trailing hemispheres. Voyager images do not reveal dramatic global leading-trailing albedo patterns, but a detailed comparison, controlling for geological age by selecting specific regions, would be needed to properly address this question. Such a study is beyond the scope of this paper.

Yet another mechanism which could produce leading-trailing asymmetries diminishing with planetocentric distance is impact cratering. In general, leading hemispheres suffer more collisions with debris from outside of the Uranus system (Zahnle et al. 2001, 2003), and gravitational focusing by Uranus causes the inner satellites to be cratered more frequently and with higher average impact velocities. Large impacts could dredge up cleaner ice from below the surfaces of the satellites, resulting in deeper $H_2O$ ice bands on leading hemispheres. Dust-sized impactors could sputter away ice and deliver exogenous, non-icy materials, creating the opposite spectral effect. Relatively little is known about the size distribution, composition, and net effect of smaller impactors on icy surfaces in the Uranus system, but the dust counter on NASA's New Horizons[1] spacecraft should provide useful information about the dust population in the ecliptic plane at large heliocentric distances.

Oberon is the most distant major satellite. It spends part of its time outside of the uranian magnetosphere in the solar wind environment (Cheng et al. 1991). This difference between Oberon and the other uranian satel-lites offers a possible explanation for its H$_2$O ice bands following a different longitudinal pattern. Closer in, magnetospheric plasma may be the dominant remover of H$_2$O ice, chiefly from trailing hemispheres, while further out, micrometeoroid impacts may increase in relative importance, preferentially sputtering ice from leading hemispheres.

---

1. NASA's first New Frontiers mission, bound for Pluto and the Kuiper belt (Stern and Cheng 2002, Stern and Spencer 2003).

## 3.2. Carbon dioxide ice

Three narrow CO$_2$ ice absorption bands are seen in some of the spectra around 2 μm. They are most apparent in Ariel spectra, where they were previously noted Grundy et al. (2003). Here we report that these bands also appear, although less prominently, in spectra of Umbriel and Titania. Figure 3 shows the CO$_2$ ice absorptions, and a general trend of more CO$_2$ ice absorption on the satellites closer to Uranus (toward the bottom of the figure): no CO$_2$ absorption is evident in Oberon's spectrum, the strongest CO$_2$ absorptions are barely visible in Titania spectra, and the absorptions become progressively stronger in spectra of Umbriel and Ariel. Two additional CO$_2$ ice absorptions at 1.578 and 1.610 μm also appear to be present in spectra of Ariel and possibly Umbriel and Titania. These absorptions are attributable to $2\nu_1 + 2\nu_2 + \nu_3$ and $\nu_1 + 4\nu_2 + \nu_3$ transitions in CO$_2$ ice, respectively (Quirico and Schmitt 1997; Gerakines et al. 2005). The $2\nu_1 + 2\nu_2 + \nu_3$ band looks like it is superimposed on the longer wavelength shoulder of a broader dip. That feature appears to be a stellar absorption that is slightly deeper in the sun than in our solar analogues, resulting in incomplete cancellation. Additional, very weak, unidentified CO$_2$ ice absorptions from the laboratory spectra of Hansen (2005) are also visible in the regions enlarged in Fig. 3. In general, SpeX does not quite spectrally resolve the cores of these CO$_2$ ice absorptions.

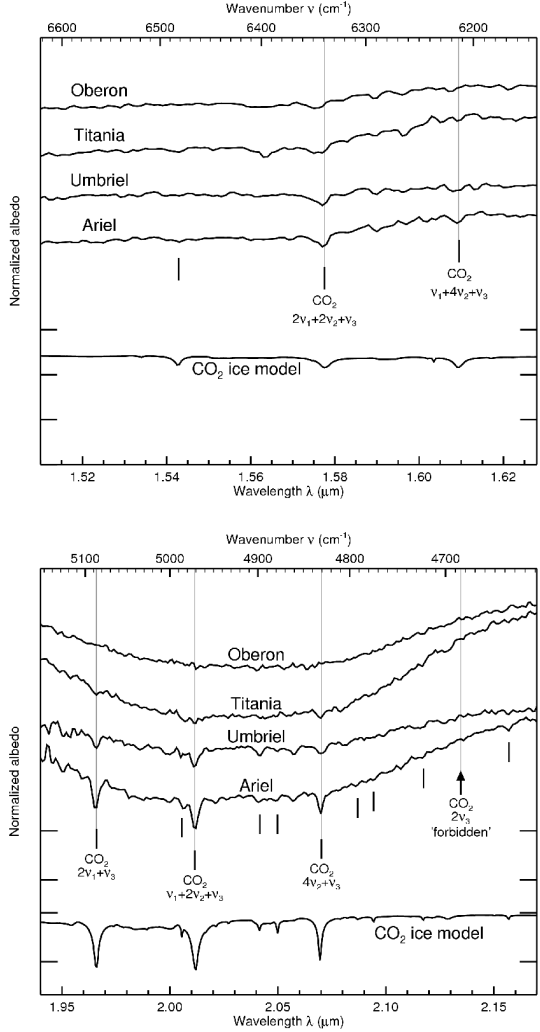

**Figure 3:** Enlarged views of the 1.6 and 2 μm regions of the same satellite spectra as in Fig. 1, showing narrow CO$_2$ ice absorption features superimposed on the broad H$_2$O ice bands. Wavelengths of additional weak, unidentified CO$_2$ ice absorptions in the laboratory data of Hansen (1997, 2005) are indicated by vertical line segments. An arbitrarily scaled Hapke model spectrum (e.g., Hapke 1993) based on those laboratory data is included at the bottom of each panel to give an idea of the sorts of band shapes expected for pure CO$_2$ ice, albeit at a warmer temperature (150 K). An arrow marks the location of the 'forbidden' $2\nu_3$ transition (Bernstein et al. 2005).

Our spectra show no apparent absorption features at 2.134 μm, where the 'forbidden' $2\nu_3$ overtone of CO$_2$ is reported by Bernstein et al. (2005) for CO$_2$ molecules mixed into H$_2$O ice (this behavior was also noted by B. Schmitt, personal communication 2002). The absence of this band couples with the wavelengths and the narrow profiles of the other CO$_2$ bands to strongly favor pure CO$_2$ ice, in the sense of CO$_2$ molecules being associated with their own kind, as opposed to being isolated in H$_2$O ice. That the CO$_2$ molecules are not isolated in H$_2$O ice is perhaps not surprising. There is some uncer-tainty about how long such a thermodynamically unstable situation would last. Laboratory evidence suggests that, at least at higher temperatures, $H_2O:CO_2$ ices mixed at the molecular level can reorganize themselves over time into segregated zones of pure phases, and thus lose the $2\nu_3$ absorption feature (Bernstein et al. 2005). It is unclear how rapidly such a reorganization could progress at the uranian satellites' surface temperatures, but the apparent ability of $H_2O$ to maintain its crystal structure against amorphization hints that such a phenomenon might not be out of the question.

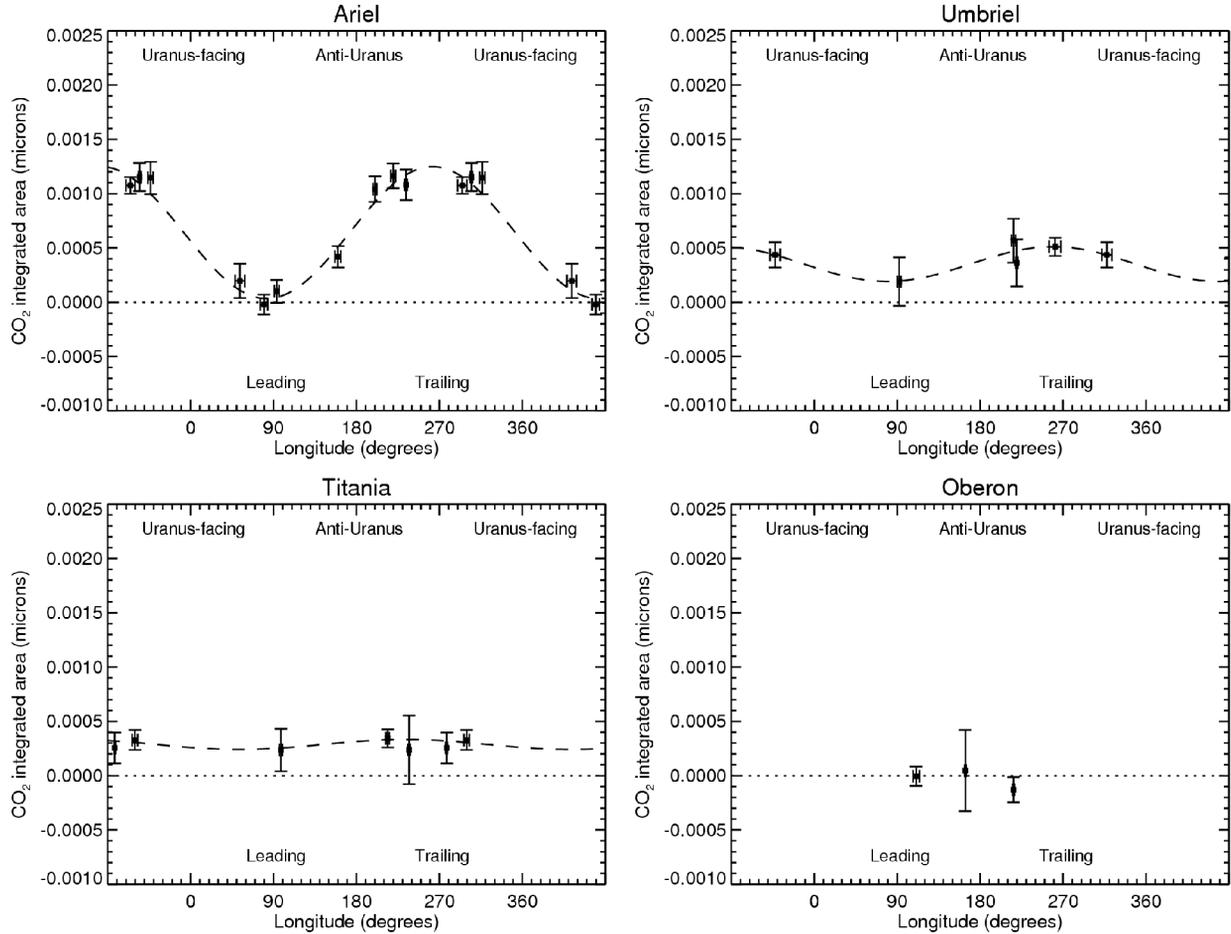

**Figure 4:** Variation of total integrated area of the three strongest $CO_2$ ice bands near 2 μm. Sine fits are plotted as dashed curves. Both the integrated areas and their longitudinal variation decline with planetocentric distance. No $CO_2$ absorption was detected in the Oberon spectra.

The spatial distribution of $CO_2$ ice can place valuable constraints on its origin and evolution. Fig. 4 explores the longitudinal and planetocentric distribution of $CO_2$ ice on the four satellites by plotting the sums of the integrated areas of the three strongest $CO_2$ ice absorptions near 2 μm from each spectrum versus sub-observer longitude. The integrated areas for the three absorptions were computed using 1.957 to 1.962 μm and 1.969 to 1.974 μm for the continuum and 1.962 to 1.969 μm for the band interval of the $2\nu_1 + \nu_3$ $CO_2$ band. For the $\nu_1 + 2\nu_2 + \nu_3$ band we used 2.002 to 2.008 μm and 2.015 to 2.020 μm for the continuum and 2.008 to 2.015 μm for the band interval. For the $4\nu_2 + \nu_3$ band we used 2.062 to 2.068 μm and 2.072 to 2.078 μm for the continuum and 2.068 to 2.072 μm for the band interval.

These measurements show more $CO_2$ ice absorption on the trailing hemispheres of Ariel, Umbriel, and Titania, as well as declining $CO_2$ absorption with distance from Uranus. This distribution contrasts somewhat with reports from Galileo/NIMS observations of Europa, Ganymede, and Callisto (McCord et al. 1997, 1998; Hibbitts et al. 2000, 2003). Those observations of the 4.25 µm $\nu_3$ fundamental band showed $CO_2$ to be present on both leading and trailing hemispheres, although more abundant at the center of the trailing hemisphere of Callisto. Also, the $\nu_3$ band is shifted in wavelength from where it occurs in pure $CO_2$ ice, suggesting that in the jovian system $CO_2$ is trapped or bound up in some other material. The absorption was also observed to be geographically associated with darker materials (Hibbitts et al. 2000, 2003). Pure $CO_2$ ice in the jovian system would not be thermally stable, adding weight to the interpretation that it is complexed with some other, less volatile material. The planetocentric trend was the opposite of what we see in the uranian system: the 4.25 µm band depth increases with distance from Jupiter (McCord et al. 1998). The $\nu_3$ band of carbon dioxide has also recently been detected in saturnian satellite spectra by Cassini/VIMS (e.g., Buratti et al. 2005; Clark et al. 2005). As in the jovian system, the band appears to be shifted from the wavelength of the pure $CO_2$ ice transition, and may be associated with dark material there, as well.

It is important to realize that the 4.25 µm $\nu_3$ fundamental absorption detected by NIMS and VIMS is intrinsically much stronger than the overtone and combination bands we are observing near 2 µm, by a factor of a thousand in absorption coefficients (e.g., Hansen 1997, 2005). Accordingly, our detection of much the weaker bands requires considerably more condensed $CO_2$ than is indicated by the jovian and saturnian satellite observations. Simple radiative transfer models (e.g., Hapke 1993; Grundy et al. 2003) can be used to get an idea of the quantities of $CO_2$ ice required by our observations. A $CO_2$ ice glaze uniformly covering Ariel's entire trailing hemisphere would have to be at least 5 µm thick to match the observed absorption bands. It would need to be thicker if the $CO_2$ ice were more geographically restricted. If the $CO_2$ ice were as segregated as possible from other materials, in a checkerboard-like pattern, at least 5% of the trailing hemisphere of Ariel would have to be composed of $CO_2$ ice.

In the following sub-sections, we consider possible sources and sinks of $CO_2$ on the satellites' surfaces, estimating rates where possible, and considering what longitudinal or planetocentric trends various processes might lead to in light of the observed $CO_2$ ice distribution. We begin with processes which can mobilize or eliminate $CO_2$ ice.

### 3.2.1. Processes which move or destroy $CO_2$

If the $CO_2$ ice on the satellite surfaces is not trapped in a less volatile material such as $H_2O$ ice or a carbonaceous residue (as is suggested by the non-detection of the $2\nu_3$ band and by the absence of wavelength shifts relative to pure $CO_2$ ice measured in the laboratory), we can use its known vapor pressure curve (Brown and Ziegler 1980) and sticking coefficient (Weida et al. 1996) to consider its stability against sublimation over the course of a seasonal cycle in the uranian system. We used a thermophysical model (e.g., Spencer and Moore 1992) to compute diurnal temperature histories as functions of latitude and season on the satellites, assuming plausible values of thermal inertia $\Gamma$, bolometric bond albedo $A_B$, and emissivity $\epsilon$. From the computed temperature histories and the $CO_2$ ice vapor pressure curve, we computed sublimation histories for each latitude band. Integrating these histories over time, we arrived at seasonally averaged sublimation rates as a function of latitude. Inverting these rates gives the time to sublimate one gram of $CO_2$ ice per square centimeter. Examples are shown in Fig. 5.

The $CO_2$ sublimation rate is proportional to its vapor pressure, which is an extremely steep function of temperature at uranian satellite surface temperatures, so local seasonal sublimation rates tend to be dominated by the seasonal maximum temperatures achieved. Polar regions, which are bathed in decades of continuous sunlight each summer because of the extreme obliquity (98°), experience the highest peak temperatures and most rapid seasonal sublimation rates. This latitude dependence of seasonal sublimation rates will tend to drive accumulation of $CO_2$ at equatorial latitudes. But even at the satellites' equators, appreciable quantities of $CO_2$ ice could sublimate on time scales much shorter than the age of the Solar System. Of the four satellites, Ariel spins fastest on its axis, with a 2.5 day period. Its faster rotation reduces the day-night temperature contrast, so for any set of thermophysical parameters, Ariel's equatorial diurnal peak temperatures remain lower than on Umbriel, Titania, and Oberon, with their longer periods of 4.1, 8.7, and 13.5 days, respectively. As a result, $CO_2$ ice is most stable against sublimation at low latitudes on the faster rotating, inner satellites.

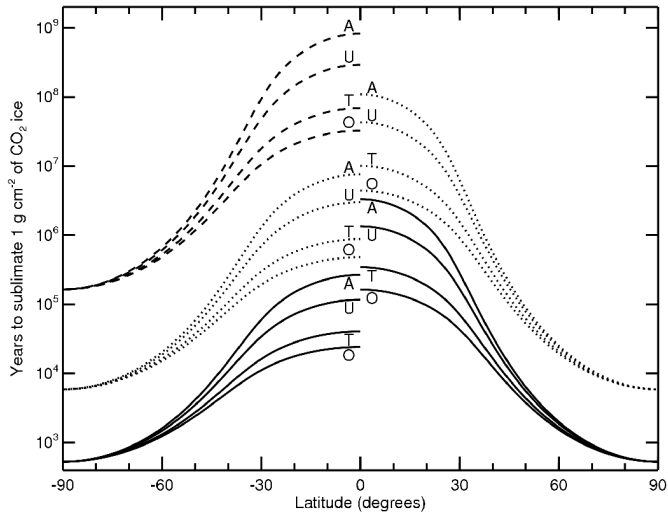

**Figure 5:** Estimates of time in Earth years to sublimate one gram of $CO_2$ per $cm^2$ as a function of latitude for satellites as labeled by their initials. North and South latitudes mirror each other, so half of each curve is omitted to reduce clutter. On the left are models with $\Gamma = 10^4$ erg $cm^{-2}$ $s^{-1/2}$ $K^{-1}$. Models with $\Gamma = 2 \times 10^4$ erg $cm^{-2}$ $s^{-1/2}$ $K^{-1}$ are on the right. Solid curves are for $A_B = 0.5$, dotted curves for $A_B = 0.6$, and dashed curves for $A_B = 0.7$. Emissivity $\epsilon = 0.9$ was assumed for all models.

The extremely strong temperature dependence makes sublimation rates highly sensitive to assumed model parameters. For instance, increasing $A_B$ from 0.5 to 0.6 reduces absorbed sunlight by 20%, leading to a few K drop in maximum temperatures. This small temperature drop drives an order of magnitude increase in the time scale to sublimate 1 g $cm^{-2}$ of $CO_2$ ice, as shown in Fig. 5. Although we assumed plausible values for albedos, thermal inertias, emissivities, etc., in our computations of temperature histories, the extreme temperature sensitivity makes the sublimation numbers highly uncertain. Our models produce peak temperatures in the 70 to 80 K range. Peak temperatures from Voyager 2 thermal emission measurements were as high as 85 K (Hanel et al. 1986). If $CO_2$ ice temperatures get that high, sublimation will proceed much more rapidly than we have estimated in Fig. 5. The thermal emission observed by Voyager is dominated by the darkest regions which absorb the most sunlight as well as topographic lows that absorb thermal radiation from surrounding areas. Any $CO_2$ ice in these warmest regions would be mobilized very rapidly.

The net result of thermal mobilization will be that $CO_2$ ice migrates away from polar latitudes toward equatorial latitudes on relatively short time scales, of the order of hundreds to millions of years. In the equatorial regions, the ice will tend to accumulate in local cold traps: regions of higher albedo, higher thermal inertia, higher emissivity, and/or higher elevation. There its texture will slowly evolve toward larger particles with rounded shapes, or even a compacted crust overlaying more porous ice below (e.g., Clark et al. 1983; Spencer 1987; Eluszkiewicz et al. 1998; Grundy and Stansberry 2000; Titus et al. 2001).

The above sublimation estimates do not account for the fate of $CO_2$ molecules after they are mobilized. Being unlikely to collide with other gas molecules, they will generally follow ballistic trajectories. Some will escape, while others will fall back to the surface and stick elsewhere. We estimate escaping fractions following Chamberlain and Hunten (1987, Eq. 7.1.5). As with sublimation rates, the fraction escaping is extremely sensitive to temperature, and escape rates from the warmer poles greatly exceed escape rates from the cooler equatorial belts. For the thermal parameters described above, we find maximum polar escape fractions ranging from $1 \times 10^{-9}$ (for Titania with $A_B = 0.7$) up to $7 \times 10^{-4}$ (for Umbriel with $A_B = 0.5$). For equatorial latitudes, peak temperatures are lower and so maximum escape fractions are lower as well, from $1 \times 10^{-10}$ to $3 \times 10^{-4}$ for the same two cases, with $\Gamma = 10,000$ erg cm$^{-2}$ s$^{-1/2}$ K$^{-1}$. For any particular set of thermal parameters, Umbriel has the highest escape fraction, followed by Ariel, because of their lower masses and gravitational accelerations. Since these two satellites show the most $CO_2$ ice, the escape of sublimated $CO_2$ to space does not seem to govern the planetocentric distribution of $CO_2$. Even for the highest equatorial escape rates of any of our models, thermal escape of $CO_2$ would be extremely slow. For Umbriel with $\Gamma = 10,000$ erg cm$^{-2}$ s$^{-1/2}$ K$^{-1}$ and $A_B = 0.5$, the time to sublimate 1 g cm$^{-2}$ at the equator is about $10^5$ years, of which $3 \times 10^{-4}$ g escape to space, so the time to lose 1 g cm$^{-2}$ to space is about $3 \times 10^8$ years. For other thermal model assumptions and for other satellites, the time scales are even longer, much longer than the age of the solar system in most cases.

Charged particle sputtering can mobilize $CO_2$, and will act at a much higher rate for $CO_2$ ice than it does for less volatile $H_2O$ ice, assuming the two ices are segregated from one another (e.g., Johnson et al. 1983). Using equation 3.27 from Johnson (2000), with values from Table 3.4 in that book, we estimate that for typical uranian satellite surface temperatures and magnetospheric proton energies, the sputtering rate for $CO_2$ ice should be some 50 to 100 times higher than for $H_2O$ ice. Cheng et al. (1991) estimate seasonal average sputtering rates for $H_2O$ molecules from uranian satellite surfaces in their Table IV. We multiply their $H_2O$ sputtering rates by 50 to 100 to get an idea of possible magnetospheric proton $CO_2$ sputtering rates for the uranian satellites, finding values in the 0.6 to $1.4 \times 10^{-8}$ g cm$^{-2}$ yr$^{-1}$ range for Ariel and Umbriel, dropping to 2 to $4 \times 10^{-10}$ g cm$^{-2}$ yr$^{-1}$ for Titania and as low as 0.5 to $1 \times 10^{-10}$ g cm$^{-2}$ yr$^{-1}$ for Oberon. As before, these values can be inverted to get an idea of how long it would take for one gram of $CO_2$ ice to be sputtered from a square centimeter. For Ariel and Umbriel, these time scales are in the 100 Myr range, increasing to more than a billion years for Titania and Oberon.

Sputtered molecules are ejected with more kinetic energy than thermally mobilized molecules generally have, so they can more readily escape a satellite's gravitational well. The energy spectrum of sputtered $CO_2$ molecules is uncertain, but crude estimates can be made by assuming it is similar to the energy distribution of sputtered $H_2O$ molecules, which goes approximately as $U'(E + U')^{-2}$, where $U'$ is a measure of binding energy and $E$ is the kinetic energy of ejection (e.g., Johnson 1998). The appropriate value of $U'$ for $CO_2$ ice is unknown. Assuming an $H_2O$-like $U' = 0.05$ eV (Haring et al. 1984) the fraction of molecules sputtered from the surfaces of the satellites at greater than escape velocity would be 0.85, 0.40, 0.44, 0.26, and 0.29 from Miranda through Oberon. If $U'$ is as low as the 0.009 eV value reported by Haring et al. (1984) for CO ice, the escape fractions would be 0.51, 0.11, 0.13, 0.06, and 0.07. These values are generally smaller than the Cheng et al. (1991) estimated escape fractions for $H_2O$ molecules, since $CO_2$ molecules are more massive (44 AMU as opposed to 18 AMU for $H_2O$). Combining the escape fractions with sputtering rates as estimated in the previous paragraph, we find time scales for one gram of $CO_2$ ice to be sputtered and lost to space to be in the few hundred million year range for Ariel and Umbriel, but longer than the age of the solar system for Titania and Oberon.

Sputtering will gradually remove $CO_2$ ice from the low latitude cold traps where thermal mobilization drives it, at least on the trailing hemispheres of the inner satellites, where magnetospheric sputtering rates are highest and the molecules are less tightly bound, gravitationally, because the satellites are less massive (see Table 1). The fact that $CO_2$ ice is observed to be most abundant in exactly these regions where it should be removed most rapidly suggests sputtering by magnetospheric protons is not controlling the distribution of $CO_2$ ice on the satellites' surfaces, and that whatever is controlling that distribution acts on shorter time scales.

Sputtering by solar UV photons could also mobilize $CO_2$ on satellite surfaces (e.g., Johnson 1990). UV

sputtering should affect all longitudes and all uranian satellites equivalently, so although it may be important for increasing mobility of $CO_2$ molecules, it cannot explain the observed spatial distribution of $CO_2$ ice. Another source of energy which can lead to sputtering is micrometeorite impacts, which, if they come from outside of the Uranus system, preferentially strike the satellites' leading hemispheres (Zahnle et al. 2003). The abundance of dust-sized impactors at 20 AU is not well known, but could be sufficient to play a role in mobilizing $CO_2$ ice from the leading hemispheres (e.g., Eviatar and Richardson 1986).

Finally, $CO_2$ molecules can be destroyed by energetic UV or charged particle radiation. Gerakines and Moore (2001) estimated $G$ value of 1.1 and 8.1 $CO_2$ molecules destroyed per 100 eV incident as protons or UV photons, respectively, via the reaction path energy + $CO_2 \rightarrow CO + O$. A UV photon with wavelength shorter than 2275 Å can dissociate a $CO_2$ molecule (Gerakines et al. 1996; Delitsky and Lane 1998). From SORCE spectra (e.g., Rottman et al. 2004), we estimate about 0.07% of the solar flux to be at or below that wavelength. Accounting for the geometry of the Uranus system, we obtain seasonal average UV fluxes ranging from 0.5 through 0.8 erg cm$^{-2}$ s$^{-1}$ for equatorial through polar latitudes on the uranian satellites. These fluxes correspond to destruction of the order of $10^{18}$ $CO_2$ molecules, or ~10 g cm$^{-2}$ yr$^{-1}$, if all of the UV photons are absorbed within the $CO_2$ ice. This is a very high destruction rate, compared with other processes we have considered.

Magnetospheric charged particles can also dissociate $CO_2$ molecules. Using the (Gerakines and Moore 2001) $G$ value of 1.1 molecule per 100 eV, along with dose rates from Cheng et al. (1991, Figure 23), we estimate radiolytic destruction rates in the $10^{-10}$ to $10^{-7}$ g cm$^{-2}$ yr$^{-1}$ range for Oberon through Miranda. These rates are much lower than the photolytic destruction rates estimated in the previous paragraph, but are comparable to the magnetospheric sputtering rates estimated previously.

Of course, not all CO and O resulting from photolysis and radiolysis of $CO_2$ will be lost. Some will recombine to make new $CO_2$ molecules while others will react with water or other locally available materials to form new molecules, some of which could be quite complex (e.g., Benit et al. 1988; Delitsky and Lane 1997, 1998). Indeed, it seems likely that a carbon cycle much like the one envisioned by Johnson et al. (2005) for the jovian system also operates on the surfaces of the uranian satellites, with carbon atoms repeatedly recycled between $CO_2$ and other molecules such as carbonic acid ($H_2CO_3$) and formaldehyde ($H_2CO$).

The relatively high destruction and loss rates estimated in this section, especially from UV photolysis as well as from sputtering and radiolysis, suggest that there must be a recent, or more likely an ongoing source for the observed $CO_2$ ice. Accordingly, we next turn our attention to possible sources of $CO_2$.

### 3.2.2. Possible sources of $CO_2$

Out-gassing from satellite interiors is a possible source for $CO_2$. Ariel, with its relatively deep $CO_2$ absorption bands, also shows clear evidence of past geologic activity (Plescia 1987), but Umbriel, with the next strongest $CO_2$ bands, shows no evidence for recent activity, at least not on the hemisphere observed by Voyager (Plescia 1989). Titania, with still less observed $CO_2$, may have some relatively youthful provinces, but its surface is mostly ancient (Zahnle et al. 2003). Although Voyager was unable to view more than about half of each satellite's surface, the presence of $CO_2$ ice does not seem to correlate particularly well with the existence of less-cratered, younger regions or with evidence of recent geological activity. Either could be an indicator of ongoing or recent release of volatiles from the satellites' interiors. An out-gassing source for the observed $CO_2$ ice thus does not look especially promising.

Impactors could deliver $CO_2$ to the uranian satellites, or could release it from their sub-surfaces. If exogenic delivery or impact exhumation were dominated by frequent, smaller impactors, they would tend to distribute $CO_2$ through the Uranus system according to a consistent pattern, and the balance between delivery and loss rates would govern the observed planetocentric and longitudinal distributions. Infrequent, larger impactors could produce a more random distribution pattern, in which a local high abundance of $CO_2$ might simply point to a single recent impact event. Zahnle et al. (2003) estimated cometary cratering rates on the uranian satellites. Owing to gravitational focusing, cratering rates per unit area increase with proximity to Uranus, being something like a factor of 4 or 5 higher on Ariel than on Oberon, and impactors strike with average speeds a factor of almost 2 higher on Ariel than on Oberon (Zahnle et al. 2003). Impactors from outside the Uranus system would preferentially strike the satellites' leading hemispheres, so the observation that $CO_2$ ice is predominantly on the trailing hemispheres would require mobilization and redistribution, followed by preferential removal from the leading hemispheres, possibly by dust impact driven sputtering.

$CO_2$ molecules could form through radiolytic action on the satellites' surfaces, with oxygen coming from the ubiquitous $H_2O$ ice, and carbon arriving via implantation of magnetospheric ions (Strazzulla et al. 2003). A magnetospheric source of carbon ions would neatly explain the observed distribution of $CO_2$ ice. However, the Voyager 2 spacecraft did not detect carbon ions in the magnetosphere during its Uranus flyby in 1986 (Bridge et al. 1986; Krimigis et al. 1986), making a magnetospheric source of carbon ions seems improbable. However, magnetospheric delivery of larger carbonaceous particles from the uranian rings remains a possibility, as mentioned previously.

There is also no shortage of dark and presumably carbonaceous material already on the satellites' surfaces, as evidenced by their low visible albedos. Recent investigations of radiolysis at interfaces between carbonaceous materials and $H_2O$ ice (Mennella et al. 2004; Gomis and Strazzulla 2005) show that $CO_2$ can be produced from purely local materials, even when the carbon and oxygen are not initially present in a single phase. The low albedos of the satellites' surfaces, along with temperatures too cold for thermal segregation of $H_2O$ ice as well as abundant craters indicative of impact gardening, suggest relatively intimate mixing between $H_2O$ ice and dark particles. In one experiment by Gomis and Strazzulla (2005), $H_2O$ ice deposited on a carbonaceous substrate was irradiated at 80 K with 200 keV $Ar^+$ ions. $CO_2$ was efficiently produced at the interface. It would be interesting to see if proton, electron, or even UV irradiation produces similar results. Any energy source capable of breaking bonds in $H_2O$ ice or in the dark carbonaceous materials, and mobilizing radicals containing O or C should be able to contribute to this process.

Whether or not radiolysis can produce $CO_2$ rapidly enough to replace $CO_2$ lost to UV photolysis is an important question to consider. Prior to saturation in the Gomis and Strazzulla (2005) experiment, each 200 keV Ar ion produced about 30 $CO_2$ molecules, or 7 keV per $CO_2$ molecule produced. This is orders of magnitude more energy than the ~12 eV per $CO_2$ molecule destroyed according to the photolytic $G$ value from Gerakines and Moore (2001). From these numbers, radiolytic production rates would be far below photolytic destruction rates. For production not to be overwhelmed by photolytic destruction, much more efficient production pathways are necessary. It is possible that lower energy protons or electrons are more effective at producing $CO_2$ from the available precursors, and it is also possible that the expected granular texture of the ice and carbonaceous material enhances production of $CO_2$, because of its large surface area. Additionally, photolysis may not be a purely destructive process. We are not aware of laboratory studies of photolytic production of $CO_2$ from $H_2O$ ice and carbonaceous precursors at uranian satellite surface temperatures, but it probably can occur (e.g., Bernstein et al. 1995; Wu et al. 2002). Photolytic production would not, however, be able to explain the observed longitudinal distribution of $CO_2$ ice. Another possibility is that more of the satellites' surfaces are involved in production than in destruction. Such a configuration is plausible, since $CO_2$ will tend to accumulate in localized cold traps, owing to its high thermal mobility, unlike the global distribution of the non-volatile pre-cursors: $H_2O$ ice and carbonaceous solids. As noted previously, the spectra are consistent with $CO_2$ being confined to a relatively small fraction of the surface area, as little as 5%.

### 3.2.3. Discussion

The planetocentric and longitudinal distributions of $CO_2$ and $H_2O$ ices in the uranian system offer valuable clues regarding their origin and fate. Having considered candidate sources and sinks on the satellite surfaces, we now review the trends expected from these mechanisms and compare them with the observational data, which shows $CO_2$ ice to be most abundant on the inner satellites, and especially on their trailing hemispheres, and $H_2O$ ice to be more abundant on leading hemispheres from Ariel through Titania, but on the trailing hemisphere of Oberon.

Sublimation mobilizes $CO_2$ relatively rapidly at seasonal maximum uranian satellite surface temperatures, which are higher at polar latitudes, so $CO_2$ accumulates at low latitudes. Relatively little is lost to space. At low latitudes, sublimation rates are much lower, but still rapid enough to drive $CO_2$ ice to accumulate in local cold traps, and to evolve in texture. Sublimation does not depend on longitude, and has no real effect on the much less volatile $H_2O$ ice. Voyager images of Umbriel show a bright equatorial feature ('Wunda') near the center of the trailing hemisphere. It is tempting to speculate that $CO_2$ ice is concentrated there, however, similar low-latitude, bright features are not evident on the trailing hemispheres of Ariel or Titania.

Impactors preferentially strike leading hemispheres and inner satellites. Impacts could dredge up fresh $H_2O$ ice from the subsurfaces, primarily of the leading hemispheres of the inner satellites, consistent with the observations of deeper $H_2O$ ice bands there. An impact source of $CO_2$ could be consistent with the observed planetocentric trend, but not with the observed longitudinal trends. An impact loss mechanism could explain the observed $CO_2$ longitudinal trends, but not the planetocentric trend. Impacts apparently cannot explain the observed $CO_2$ ice distribution, at least not without the help of additional mechanisms.

Magnetospheric charged particle sputtering preferentially mobilizes $CO_2$ and $H_2O$ ices from the trailing hemispheres of the inner satellites, exactly those regions where $CO_2$ ice is observed. Sputtering rates are estimated to be well below sublimation rates for $CO_2$, but are probably the dominant sink for $H_2O$ ice. A much greater fraction of sputtered $CO_2$ molecules should escape to space, compared with sublimated $CO_2$ molecules. Correlations with indicators of recent geological activity in Voyager images are not evident, arguing against an out-gassing source of $CO_2$ from the satellite interiors.

Radiolytic production of $CO_2$ driven by magnetospheric irradiation should act most rapidly on the trailing hemispheres of the inner satellites, because magnetospheric densities are higher closer to Uranus. The longitudinal and planetocentric pattern of radiolytic production is consistent with the observed distribution of $CO_2$, and appears to be the best candidate source for $CO_2$ on the satellites' surfaces. The fact that our estimated rates of radiolytic production are considerably below the estimated photolytic destruction rate is troubling. This discrepancy may point to problems in our estimated production rates or could imply that much more surface area is involved in production than is actually covered with $CO_2$ ice, in order to maintain an equilibrium. The idea that $CO_2$ ice may only cover a small fraction of the surface area of the satellites' trailing himespheres is at least consistent with the spectral evidence.

Finally, it seems probable that $CO_2$ ice on the uranian satellites' surfaces participates in a radiolytic and photolytic carbon cycle (e.g., Johnson et al. 2005). If so, carbonic acid is also likely to be present, and might be detectable by means of its characteristic infrared absorption bands. With hydrogen being more readily lost to space than heavier carbon and oxygen, chemistry at the satellites' surfaces probably tends to be oxidizing and acidic. Such a surface environment may help explain why $NH_3$, an alkali, has not been convincingly detected, despite its presence being suggested by geomorphologic evidence (Croft and Soderblom 1991; Kargel et al. 1991). A tentative report of $NH_3$ in a spectrum of Miranda (Bauer et al. 2002) has yet to be confirmed. If $NH_3$ were confirmed to be present on the surface of a uranian satellite, the chemical implications would be quite remarkable.

## 4. Conclusion

Seventeen nights of IRTF/SpeX observations of the 4 largest uranian satellites reveal the existence of $CO_2$ ice on Umbriel and Titania for the first time, as well as confirming its presence on Ariel. The observations also provide information about the longitudinal and planetocentric distributions of $CO_2$ and $H_2O$ ices. These distributions likely result from equilibria between source and sink mechanisms, and can tell us something about the relative importance of the processes involved. Magnetospheric sputtering removal of $H_2O$ ice from trailing hemispheres of Ariel, Umbriel, and Titania is consistent with the observation of deeper $H_2O$ ice absorption bands on the leading hemispheres of those satellites, with diminishing leading-trailing asymmetries with planetocentric distance. For Oberon, magnetospheric sputtering is greatly reduced, and the leading-trailing asymmetry appears to be reversed. The deeper $H_2O$ ice bands on Oberon's trailing hemisphere may point to an effect of micrometeoroid sputtering that only becomes important where magnetospheric effects become negligible. Radiolytic production of $CO_2$, driven by magnetospheric charged particle bombardment, appears most consistent with the observation of deeper $CO_2$ ice absorption bands on the trailing hemispheres of Ariel, Umbriel, and Titania, as well as declining $CO_2$ absorption with planetocentric distance and its absence from the surface of Oberon.

Thermally-driven processes such as grain growth, volatile transport, sintering, and solar gardening can be expected to influence the texture and distribution of $CO_2$ ice on the surfaces of the uranian satellites. The seasonally-averaged distribution of sunshine should drive $CO_2$ ice away from polar latitudes. Low latitude regions which radiatively cool fastest will cold-trap $CO_2$ ice, and if enough ice accumulates to raise the albedos of these regions, the process will be self-reinforcing. Uppermost surfaces of $CO_2$ ice particles can radiatively cool faster than subsurface depths where solar energy is absorbed, so sublimation will tend to preferentially remove $CO_2$ from below the uppermost surface. Loss to space driven primarily by sputtering by UV photons and magnetospheric charged particles limits the residency time of $CO_2$ ice in these cold traps. Additional limits on $CO_2$ accumulation come from UV photolysis, which is expected to break $CO_2$ bonds at a relatively high rate, leading to a chemical cycle in which carbon, oxygen, and hydrogen are repeatedly combined into new species and broken apart again, forming a complex mixture which should include $CO_2$ as well as species like carbonic acid and formaldehyde. More detailed comparisons between leading hemispheres (which see relatively little magnetospheric radiolysis) and trailing hemispheres should be able to shed more light on the relative importance of radiolysis and photolysis.

If radiolysis by magnetospheric particles is indeed the source of $CO_2$ ice on the uranian satellites, production rates on Miranda should be even higher than on Ariel. However, Miranda's lower gravity will allow a much larger fraction of sublimated and sputtered $CO_2$ molecules to escape. Spectral observations of Miranda could provide useful constraints on the balance between sublimation and radiolytic production.

ACKNOWLEDGMENTS: We are grateful to W. Golisch, D. Griep, P. Sears, S.J. Bus, J.T. Rayner, and K. Crane for assistance with the IRTF and with SpeX, R.S. Bussmann for contributing to the reduction pipeline, M.R. Showalter for the Rings Node's on-line ephemeris services, and NASA for its support of the IRTF. We especially thank K.A. Tryka for sharing with us an unpublished manuscript on $NH_3$ and $CO_2$ stability in the uranian system, which contained many useful ideas for modeling the fate of $CO_2$ ice on the satellites' surfaces. Scientific discussions with B. Schmitt and M.P. Bernstein also provided useful ideas. We thank the free and open source software communities for empowering us with many of the tools used to complete this project, notably Linux, the GNU tools, LATEX, FVWM, Tcl/Tk, TkRat, and MySQL. We acknowledge the significant cultural role and reverence for the summit of Mauna Kea within the indigenous Hawaiian community and are honored to have had the opportunity to observe there. Finally, we are grateful for funding from NSF grants AST-0407214 and AST-0085614 and from NASA grants NAG5-4210, NAG5-10497, NAG5-12516, and NNG04G172G, without which this work could not have been done.